\documentclass[journal]{IEEEtran}

\usepackage{amsthm,amsmath,amssymb}
\usepackage{eucal}
\usepackage{xifthen}
\usepackage{mathtools}
\usepackage{enumerate}
\usepackage{microtype}
\usepackage{xspace}
\usepackage{bm}
\usepackage{cite}
\usepackage{fancyhdr}
\usepackage{lastpage}
\usepackage[]{caption}
\usepackage{xcolor}
\allowdisplaybreaks

\long\def\comment#1{}

\newfont{\bbb}{msbm10 scaled 700}

\newfont{\bb}{msbm10 scaled 1100}

\newcommand{\RR}{\mbox{\bb R}}

\newcommand{\mbs}[1]{\bm{#1}}

\newcommand{\mat}[1]{{\uppercase{\mbs{#1}}}}

\newcommand{\Pmatrix}[1]{\begin{array}{ll}#1\end{array}}

\renewcommand{\Re}[1][]{\ifthenelse{\isempty{#1}}{\operatorname{Re}}{\operatorname{Re}\left(#1\right)}}
\renewcommand{\Im}[1][]{\ifthenelse{\isempty{#1}}{\operatorname{Im}}{\operatorname{Im}\left(#1\right)}}

\newcommand{\Am}{\mat{a}}

\newcommand{\Tm}{\mat{t}}

\newcommand{\Cc}{{\mathcal C}}
\newcommand{\Dc}{{\mathcal D}}
\newcommand{\Ec}{{\mathcal E}}

\newcommand{\Gc}{{\mathcal G}}
\newcommand{\Hc}{{\mathcal H}}

\newcommand{\Mc}{{\mathcal M}}

\newcommand{\Qc}{{\mathcal Q}}

\newcommand{\Sc}{{\mathcal S}}

\newcommand{\Wc}{{\mathcal W}}
\newcommand{\Vc}{{\mathcal V}}

\newcommand{\CN}[1][]{\ifthenelse{\isempty{#1}}{\mathcal{N}_{\mathbb{C}}}{\mathcal{N}_{\mathbb{C}}\left(#1\right)}}
\renewcommand{\P}[1][]{\ifthenelse{\isempty{#1}}{\mathbb{P}}{\mathbb{P}\left(#1\right)}}
\newcommand{\E}[1][]{\ifthenelse{\isempty{#1}}{\mathbb{E}}{\mathbb{E}\left(#1\right)}}
\renewcommand{\det}[1][]{\ifthenelse{\isempty{#1}}{\mathrm{det}}{\mathrm{det}\left(#1\right)}}
\newcommand{\trace}[1][]{\ifthenelse{\isempty{#1}}{\mathrm{tr}}{\mathrm{tr}\left(#1\right)}}
\newcommand{\rank}[1][]{\ifthenelse{\isempty{#1}}{\mathrm{rank}}{\mathrm{rank}\left(#1\right)}}
\newcommand{\diag}[1][]{\ifthenelse{\isempty{#1}}{\mathrm{diag}}{\mathrm{diag}\left(#1\right)}}

\DeclarePairedDelimiter\abs{\lvert}{\rvert}

\renewcommand{\Re}{{\rm Re}}
\renewcommand{\Im}{{\rm Im}}

\newtheorem{remark}{Remark}
\newtheorem{definition}{Definition}
\newtheorem{theorem}{Theorem}
\newtheorem{example}{Example}
\newtheorem{lemma}{Lemma}

\usepackage{hyperref}
\hypersetup{
    bookmarks=true,         
    unicode=false,          
    pdftoolbar=true,        
    pdfmenubar=true,        
    pdffitwindow=false,     
    pdfstartview={FitH},    
    pdfnewwindow=true,      
    colorlinks=true,       
    linkcolor=red,          
    citecolor=green,        
    filecolor=magenta,      
    urlcolor=cyan           
}

\DeclareMathAlphabet{\mathcal}{OMS}{cmsy}{m}{n}

\usepackage{graphicx}
\usepackage[font=small]{caption}
\usepackage{subcaption}
\usepackage{mathrsfs}

\begin{document}
\title{TDMA  is Optimal for  All-unicast DoF Region of TIM 
if and only if Topology is Chordal Bipartite
}

\author{\IEEEauthorblockN{Xinping Yi, Hua Sun, Syed A. Jafar, and David Gesbert}
\thanks{X.~Yi is with Communications and Information Theory group at Technical University of Berlin, 10587 Berlin, Germany. (email: {\tt xinping.yi@tu-berlin.de}). H.~Sun and S.~Jafar are with CPCC at University of California Irvine, 92697 Irvine, CA, USA (email: {\tt \{huas2,syed\}@uci.edu}). D.~Gesbert is with EURECOM, 06410 Biot, France (email: {\tt gesbert@erurecom.fr}). The work of X. Yi and D. Gesbert was carried out within the framework of Celtic-Plus SHARING project.}
}
\maketitle


\begin{abstract}
The main result of this work is that an orthogonal access scheme such as TDMA achieves the all-unicast degrees of freedom (DoF)  region of the topological interference management (TIM) problem if and only if the  network topology graph is chordal bipartite, i.e., every cycle that can contain a chord, does contain a chord. The all-unicast DoF region includes the DoF region for any arbitrary choice of  a unicast message set, so e.g., the results of Maleki and Jafar  on the optimality of orthogonal access for the sum-DoF of one-dimensional convex networks are recovered as  a special case. The result is also established for the corresponding topological representation of the index coding problem.
\end{abstract}

\section{Introduction}
The topological interference management problem (TIM), introduced in \cite{Jafar_TIM},  studies the degrees of freedom (DoF) of partially connected one-hop wireless networks with no channel state information at the transmitters  except the network topology. As a generalization of the classical optimal frequency reuse question, and for its use of robust interference alignment schemes, TIM is of much practical interest. It is also of great theoretical interest because of an essential equivalence between TIM and the index coding problem, established in \cite{Jafar_TIM}. The index coding problem is one of the most intriguing open problems in network information theory due to its rich connections to various other prominent problems ranging from distributed storage \cite{Arya_Storage, Karthikeyan}, caching \cite{Maddah_Ali_Niesen} and general instances of network coding \cite{Rouayheb_Sprintson_Georghiades, Effros_Rouayheb_Langberg} to  hat-guessing problems  \cite{Riis_Hat} in recreational mathematics. 

On the surface these problems seem very simple. However, in spite of various graph-theoretic \cite{Yossef_Birk_Jayram_Kol_Trans}, random coding \cite{Kim_region}, rate-distortion \cite{Wagner_index}, as well as interference alignment \cite{Maleki_Cadambe_Jafar} approaches that have been brought to bear upon it, the index coding problem, and by association the TIM problem, remain open. The difficulty is evident in recent results for index coding that prove the necessity, in general, of  non-linear coding schemes for achievability arguments \cite{Rouayheb_Sprintson_Georghiades, Blasiak_Kleinberg_Lubetzky_2011, Maleki_Cadambe_Jafar}, and of non-Shannon information inequalities for converses \cite{Blasiak_nonshannon, Sun_Jafar_nonshannon, Riis_nonshannon}, neither of which are well understood. As such, much of the recent progress on these problems has come about from a divide and conquer approach aimed at identifying broad classes of solvable instances of the problem. For example,  \cite{Yossef_Birk_Jayram_Kol_Trans, Neely_Tehrani_Zhang_Trans} identify all instances where it is optimal to serve only one user at a time. All half-rate feasible instances are identified in \cite{Blasiak_nonshannon, Maleki_Cadambe_Jafar}.  All instances for which the alignment graph has no cycles or no forks are solved in \cite{Jafar_TIM} and this class is further generalized to include instances for which the alignment graph has no overlapping cycles in \cite{Sun_Jafar_nonshannon}.  All instances with 5 or fewer messages are solved in \cite{Kim_region}, all single uniprior instances are solved in \cite{Ong_single-uniprior}, and all instances where each source has non-trivial communication capacity to at most two of the destinations are solved in \cite{Wagner_index}. 

Continuing with the divide and conquer strategy, a problem of great interest from both theoretical and practical perspectives, is to identify those instances where the simplest achievable schemes are also optimal. Such an approach has been quite successful recently in wireless interference networks where much progress has been made in identifying settings where the simple scheme of treating interference as noise is (in a generalized DoF sense) optimal \cite{Geng_TIN}. For TIM, perhaps the simplest scheme is an orthogonal access scheme, such as TDMA. The corresponding scheme for index coding is graph coloring, i.e., simultaneous scheduling of only those messages that do not interfere with each other.  Remarkably, in recent work by \cite{Maleki_Jafar_Convex},   TDMA  and graph coloring are shown to achieve the sum-DoF and sum-capacity of a fairly broad class of TIM and index coding problems, respectively. Based on the constraints on the topology and message sets, this class is  known as one-dimensional convex networks. This  is our starting point.

\subsection{One-Dimensional Convex Networks \cite{Maleki_Jafar_Convex}}
\begin{figure}[htb]
\center
\includegraphics[width=0.9\columnwidth]{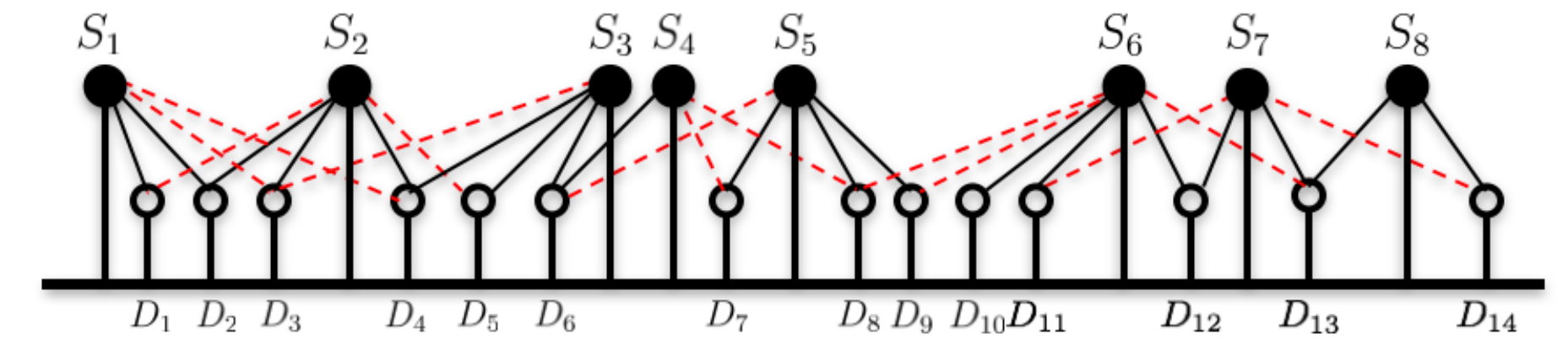}
\caption{A one-dimensional convex network instance of TIM (Fig. 2 of \cite{Maleki_Jafar_Convex}). Each solid black edge indicates an independent desired message, each dashed red edge indicates no desired message, and the absence of an edge indicates `out-of-range'. All edges together comprise the bipartite network topology graph.
}
\label{fig:convex}
\end{figure}

 For the TIM problem,  \cite{Maleki_Jafar_Convex} shows that orthogonal access (such as time division multiple access --- TDMA) achieves the sum-DoF of a one-dimensional cellular network (all nodes placed on a straight line) that satisfies (i) source convexity, (ii) destination convexity, and (iii) message convexity. An example of such a network (with 19 messages) is shown in Fig. \ref{fig:convex}. The convexity assumptions are motivated by the observation that signals are stronger and communication is more likely to occur between nodes that are physically closer to each other than between nodes that are farther apart. For example, since Source $S_1$ is heard by Destination $D_4$, it must also be heard by destinations that are closer to it than $D_4$, e.g., destinations $D_1, D_2, D_3$. This is referred to as source convexity. Destination convexity is similarly defined. Further, if Source $S_1$ has a desired message for Destination $D_2$, it must have a desired message for Destination $D_1$ because $D_1$ is closer to $S_1$ than $D_2$. This is denoted as the message convexity assumption. Under these three assumptions, \cite{Maleki_Jafar_Convex} shows that TDMA is optimal in terms of sum-DoF. For example, since the network in Fig. \ref{fig:convex} satisfies these assumptions, an orthogonal scheduling scheme that schedules messages  only between non-interfering source-destination
pairs $S_1 \rightarrow D_1, S_3 \rightarrow D_5, S_5 \rightarrow D_7, S_6 \rightarrow D_{10}$ and $S_8 \rightarrow D_{14}$, achieves the optimal sum-DoF value ($5$ in this case).

\begin{figure}[htb]
\center
\includegraphics[width=0.8\columnwidth]{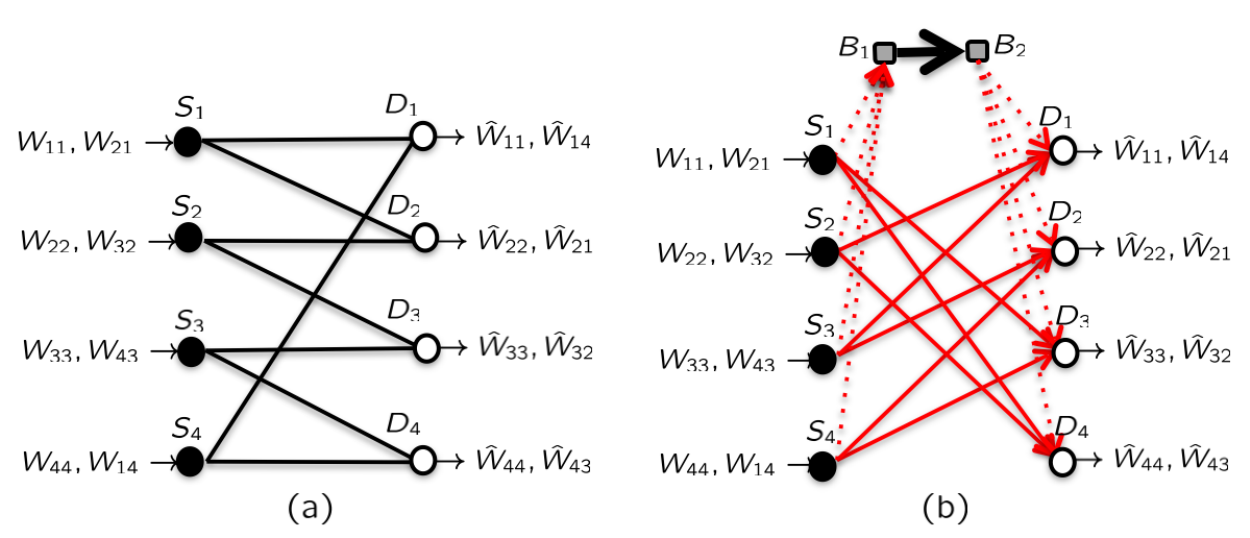}
\caption{(a) An instance of the TIM problem (edges constitute the bipartite network topology graph)  and (b) the corresponding instance of the index coding problem. Red links (solid and dashed) have infinite capacity. Solid red links form the antidote graph, which is the complement of the network topology graph.
}
\label{fig:timic}
\end{figure}

The TIM result corresponds directly to an index coding result due to the remarkable association between the two problems identified in \cite{Jafar_TIM}, such that there is a TIM instance associated with each index coding instance and vice versa (see Fig. \ref{fig:timic}). Corresponding instances of both problems are jointly described by a bipartite topology graph with sources on one side, destinations on the other, and edges representing the presence of a non-trivial communication channel whose communication capacity  is not zero (TIM) or infinity (index coding). Reference \cite{Jafar_TIM} shows that (expressed in normalized units) the capacity region of any instance of the index coding problem acts as an outer bound on the DoF region of the corresponding instance of the TIM problem, and furthermore, the two are equivalent when restricted to linear schemes over the same field. 

\subsection{Beyond One-Dimensional Convex Networks}
On the one hand, the optimality of TDMA for the physically motivated and fairly broad class of topologies represented by one-dimensional convex networks is  surprising because it is known that simple schemes such as TDMA or CDMA (cf. (fractional) coloring and (partition) multicast) can be severely sub-optimal in general. For example, there exist instances of TIM (index coding) with $K$ messages where optimal schemes involving interference alignment achieve a sum-DoF (sum-capacity) value that is a factor of $(1/3+o(1))K^{1/4}$ higher than the best achievable through TDMA or CDMA \cite{Jafar_TIM, Blasiak_nonshannon}. 

On the other hand, however, the result is limited by the assumption of a one-dimensional placement of nodes and the convexity constraints. For instance, even for physically motivated TIM topologies that satisfy all the convexity constraints, it is shown in \cite{Maleki_Jafar_Convex} that going from one-dimensional settings to the much more realistic two-dimensional placements of sources and destinations, one immediately runs into examples where TDMA is no longer optimal\footnote{Fig. \ref{fig:timic} is such an example. It corresponds to a 2 dimensional convex network (see \cite{Jafar_TIM}) with optimal sum-DoF value of 8/3, achieved by interference alignment, whereas orthogonal schemes cannot achieve more than 2 DoF.} and interference alignment solutions significantly outperform conventional baselines. The convexity assumptions are also not applicable to heterogeneous networks, where  a user may hear a distant high power base station, but still not be able to hear a closer but lower power base station. Moreover, beyond the TIM context, for the index coding problem in general, the one-dimensional node placements or convexity constraints are of little physical significance. Last but not the least, the focus on sum-DoF (capacity) is restrictive as well. 

This brings us to the motivation of this work, which is to go beyond  these limitations, to answer the question --- \emph{what is the fundamental topological structure that determines the  optimality of TDMA (fractional coloring) for the TIM (index coding) problem,  making  other sophisticated schemes redundant?}

\subsection{Summary of Contribution}
\begin{figure}[htb]
 \centering
\center
\includegraphics[width=0.7\columnwidth]{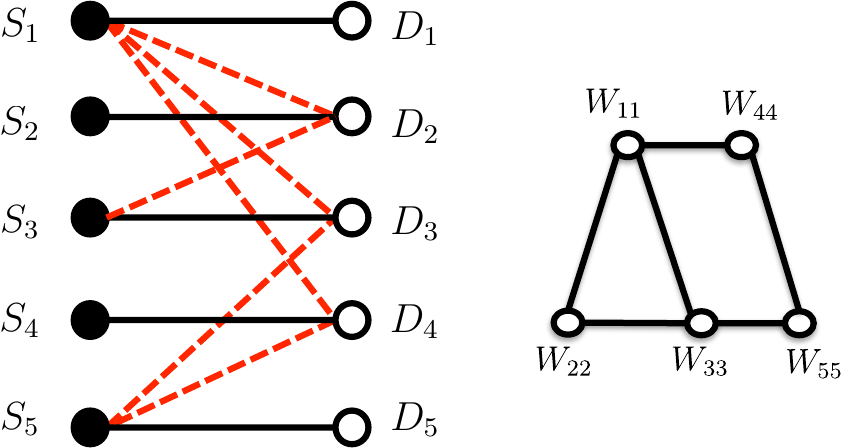}
\caption{A 5-message TIM instance (left), and  its conflict graph (right). See Section \ref{sec:model} for details.}
\label{fig:IC_ex}
\end{figure}

The main contribution of this work,\footnote{Frequently in this paper, we will use a compact notation where we merge corresponding statements for TIM and index coding by using parantheses for the respective alternatives.} as highlighted in the title, is to show that TDMA (fractional coloring) achieves the all-unicast DoF (capacity) region of the TIM (index coding) problem if and only if the network topology graph is chordal bipartite, i.e.,  any cycle that can contain a chord,  contains a chord.\footnote{Note that chords are only defined for cycles, so e.g., tree topologies are also chordal bipartite graphs. } 
 Note that  network topology graphs are always bipartite, and for such graphs, cycles of length 4 cannot have a chord. So a chordal bipartite graph is one in which any cycle of length 6 or more (such cycles are called long cycles) must contain a chord.
For example, consider the network topology graph in Fig. \ref{fig:IC_ex}, it is chordal bipartite because it only contains one long cycle (formed by nodes $S_1, D_2, S_3, D_3, S_5, D_4$), and this cycle has a chord (the edge connecting $S_1$ and $D_3$). Note that for a chordal bipartite graph, it is acceptable for a long cycle to have \emph{some} chords missing (e.g., in Fig. \ref{fig:IC_ex}, the chord connecting nodes $D_2$ and $S_5$ is not present in the only long cycle), provided it does not have \emph{all} its chords missing.
Furthermore, whether or not a network is chordal bipartite is easy to check (polynomial time) \cite{weakly-recog}.  

The all-unicast setting involves an independent message from each source to each destination, i.e., all possible unicast messages. A unicast message is one that originates at only one source and is desired by only one destination. Characterizing the all-unicast DoF (capacity) region automatically characterizes the DoF (capacity) region for any arbitrary subset of unicast messages, as well as other traditional metrics such as symmetric or sum DoF (capacity). So, when TDMA achieves the all-unicast DoF region, it achieves the DoF region of any arbitrary subset of messages as well. 
For example, consider again Fig. \ref{fig:IC_ex}. Since the network topology graph is chordal bipartite, our result implies that TDMA achieves the all-unicast DoF region and therefore also the DoF region of arbitrary subset of messages (e.g., the interference channel message setting, where each Source $S_i, i \in \{1,2, \cdots, 5\}$ has an independent message only for its corresponding Destination $D_i, i \in \{1,2,\cdots,5\}$.)

\begin{figure}[htb]
 \centering
\includegraphics[width=1\columnwidth]{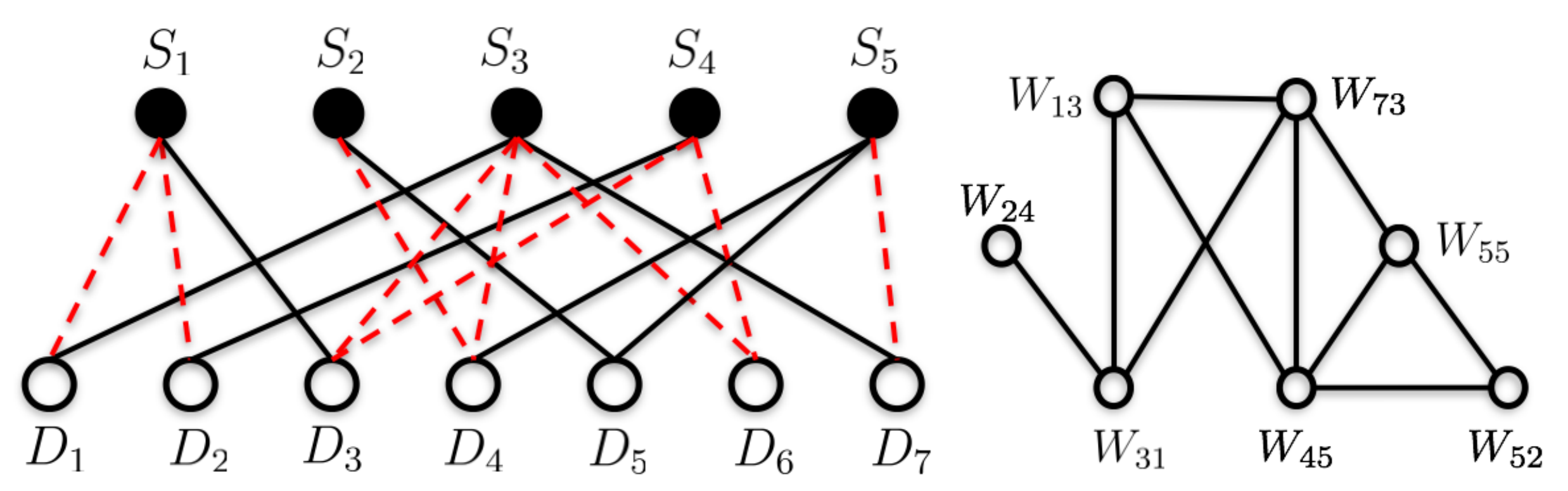}
\caption{A 7-message TIM instance (left) and its conflict graph (right). See Section \ref{sec:model} for details.
}
\label{fig:chordal_ex}
\end{figure}

While not restricted to the one-dimensional convex topologies studied in \cite{Maleki_Jafar_Convex}, chordal bipartite networks do include them as a special case. Since convex message sets are included in all-unicasts, and the DoF region includes sum-DoF, the results of \cite{Maleki_Jafar_Convex} are recovered as a special case of our result. Within the context of one-dimensional networks, our result  generalizes the results of \cite{Maleki_Jafar_Convex} --- it shows that for one-dimensional placement of sources and destinations, if the network satisfies either source convexity \emph{or} destination convexity (without requiring \emph{both}, turns out either assumption is sufficient (not necessary) to imply a chordal {bipartite} topology), then for any arbitrary unicast message set (without requiring message convexity), the entire DoF region (without restriction to only sum-DoF) is achieved by TDMA. For the setting illustrated in Fig. \ref{fig:convex} (note that the topology is chordal {bipartite}), the result shows that the entire DoF (capacity) region for the 19 messages  (as well as any other set of unicast messages possible in this setting) is achieved by TDMA (fractional coloring). Other examples that further highlight the generality of this result are presented in  Fig. \ref{fig:IC_ex} (does not satisfy destination convexity) and Fig. \ref{fig:chordal_ex} (does not satisfy source or destination convexity) where also the network topology graph is chordal {bipartite} and TDMA (fractional coloring) is optimal for the DoF (capacity) region under all possible unicast message sets. On the other hand, Fig. \ref{fig:timic} is an example where the topology is not chordal  bipartite, and so TDMA (fractional coloring) schemes cannot achieve the DoF (capacity) region (as  indeed is shown in \cite{Jafar_TIM}).

So to answer the question that motivates this work ---  \emph{chordal {bipartite} network topology is the fundamental topological structure that determines the optimality of TDMA (fractional coloring), making all other sophisticated schemes unnecessary.}

\section{System Model}\label{sec:model}
Corresponding instances of TIM and index coding problems, each with $M$ sources and $N$ destinations, are simultaneously specified by a topology matrix $\Tm$ and a message set $\Mc$.  $\Tm$ is an $N\times M$ matrix with elements $t_{ji}\in\{0,1\}$. If there exists a non-trivial channel from Source $i$ to Destination $j$, i.e., a channel whose capacity is not zero (TIM) or infinity (index coding) then $t_{ji}=1$, otherwise $t_{ji}=0$. Since communication is non-trivial only if the channel capacity is not zero or infinity, the set of  messages $\Mc$ is a subset of $\overline{\Mc}\triangleq \{W_{ji}: t_{ji}=1, i \in \{1,\dots,M\}, j \in \{1,\dots,N\}\}$, with $W_{ij}$ representing an independent unicast message that originates at Source $i$ and is intended for Destination $j$. If $\Mc=\overline{\Mc}$, the setting is called the all-unicast setting. Corresponding instances of TIM and index coding share  the same topology matrix $\Tm$ and the same message set $\Mc$. The remaining description for each problem is provided next.

\subsection{Topological Interference Management Problem (TIM)}
An arbitrary instance of the TIM problem \cite{Jafar_TIM} is represented as a partially connected network with $M$ sources, labeled $S_1$, $S_2$, $\dots$, $S_M$, and $N$ destinations, labeled $D_1$, $D_2$, $\dots$, $D_N$.  All sources and destinations are equipped with a single antenna each.
The received signal  for Destination $D_j$ at time instant $t$ is:
\begin{align}
Y_j(t) = \sum_{i = 1}^{M} t_{ji} h_{ji}(t) X_i(t) + Z_j(t)
\end{align}
where $X_i(t)$ is the transmitted signal from Source $S_i$. All transmitted signals are subject to a power constraint $P$. $Z_j(t)$ is the Gaussian noise with zero-mean and unit-variance at Destination $D_j$, $h_{ji}(t)$ is the channel coefficient between Source $S_i$ and Destination $D_j$. The topology matrix  $\Tm = [t_{ji}]_{N \times M}$ is known by all sources and destinations. Only desired channel coefficients are known by destinations.
The channel coefficients and network topology are assumed to be fixed throughout the duration of communication. Time-varying channels will be discussed in Section \ref{sec:time-varying}.

{Let us define two graphs for the TIM problem.
\begin{definition} 
\label{def:tim-topology}
Given the TIM problem with $M$ sources and $N$ destinations,  topology matrix ${\Tm}$, and message set  $\Mc \subseteq \overline{\Mc}$, define the following two graphs:
\begin{itemize}
\item {\bf Network Topology Graph}: An undirected bipartite graph with sources on one side, destinations on the other, and an edge between $S_i$ and $D_j$ whenever $t_{ji}=1$.
\item {\bf Message Conflict Graph}: An undirected graph where each message $W_{ji} \in \Mc$ is a vertex, and edges exist between two vertices if the two messages conflict with each other. Two messages $W_{ji},W_{j'i'} $ conflict if $(t_{j'i}, t_{ji'}) \neq (0,0)$, i.e., they originate from the same source ($i = i'$), or are intended for the same destination ($j = j'$), or if the source of one message interferes with the  destination of the other message.
\end{itemize}
\end{definition}
Note that the network topology graph depends only on network connectivity, regardless of the message demand.
The message conflict graph indicates the conflict between two messages, but it loses some information of conflicting source and destination.

For TIM we use DoF region as our figure of merit. Common terms such as the coding scheme, achievable rate tuple $(\bar{R}_{ji}: W_{ji} \in \Mc)$, capacity region $\Cc$ are used in the standard Shannon theoretic sense. Unfamiliar readers can refer to \cite{Jafar_TIM} for details.
\begin{definition} [DoF Region]
\begin{align}
\MoveEqLeft {\Dc}_{\Mc}=\left\{(d_{ji}: W_{ji} \in \Mc) \in \RR_+^{\abs{\Mc}}:  \right. \nonumber \\ &  \left. d_{ji} = \lim_{P \to \infty}  \frac{\bar{R}_{ji}}{\log P}, ~ {\rm s.t.} ~ (\bar{R}_{ji}: W_{ji} \in \Mc) \in {\Cc} \right\}
\end{align}
\end{definition}
}

\subsection{Index Coding Problem (IC)}

The index coding problem consists of $M$ sources, labeled $S_1, S_2, \ldots, S_M$, $N$ destinations, labeled $D_1, D_2, \ldots, D_N$ and two additional nodes $B_1, B_2$, that are connected by a unit capacity edge from $B_1$ to $B_2$. There is an infinite capacity link from every source to $B_1$, and from $B_2$ to every destination. We also have infinite capacity links between sources and destinations identified by the antidote matrix { ${\Am} = [a_{ji}]_{N \times M}$}, where $a_{ji} = 1$ means that Source $S_i$ is connected to Destination $D_j$ through an infinite capacity link, and $a_{ji} = 0$ otherwise. The antidote matrix $\Am$ is simply the complement of the topology matrix $\Tm$ and is obtained as $\Am=[1]_{N\times M}-\Tm$, where $[1]_{N\times M}$ denotes the $N\times M$ matrix with all entries equal to $1$. 

Define these graphs for the index coding problem.
\begin{definition}
\label{def:repic}
Given the index coding problem with $M$ sources and $N$ destinations,  antidote matrix ${\Am}$, and message set  $\Mc \subseteq \overline{\Mc}$, define the following graphs:
\begin{itemize}
\item {\bf Antidote Graph}: A directed bipartite graph with sources on one side, and destinations on the other, and an arc (i.e., directed edge) from $S_i$ to $D_j$ whenever $a_{ji}=1$.
\item {\bf Side Information Digraph}: A directed graph where each message $W_{ji} \in \Mc$ is a vertex, and there exists an arc (i.e., directed edge) from $W_{ji}$ to $W_{j'i'}$ if 
$a_{j'i}=1$.
\item {\bf Network Topology Graph}: An undirected bipartite graph with sources on one side, destinations on the other, and an edge between $S_i$ and $D_j$ whenever $a_{ji}=0$.
\item {\bf Message Conflict Graph}: An undirected graph where each message $W_{ji} \in \Mc$ is a vertex, and edges exist between two vertices if the two messages conflict with each other. Two messages $W_{ji},W_{j'i'} $ conflict if $(a_{j'i}, a_{ji'}) \neq (1,1)$, i.e., they originate from the same source ($i = i'$), or are intended for the same destination ($j = j'$), or if one source is not connected to the other destination through an infinite capacity link.
\end{itemize}
\end{definition}

Note that given the topology matrix $\Tm$ and the message set $\Mc$,  corresponding instances of TIM and index coding have the same network topology graph and the same message conflict graph. 


Coding schemes, achievable rate tuple $({R}_{ji}: W_{ji} \in \Mc)$ and capacity region are used in the standard Shannon theoretic sense. 

Clique covering on the side information digraph is one of the most basic coding schemes in index coding. A set of messages forms a {\em di-clique} if, for any two messages $W_{ji}$ and $W_{j'i'}$, there is an arc from $W_{ji}$ to $W_{j'i'}$ and an arc from $W_{ji}$ to $W_{j'i'}$. Clique covering is to partition the side information digraph into di-cliques such that all vertices are covered at least once. The messages in each di-clique of the side information digraph are non-adjacent in its complement, and thus non-adjacent in the message conflict graph. That is, the messages in the di-clique of the side information digraph form an independent set in the (undirected) message conflict graph. 
As such, the clique covering in the side information digraph is equivalent to vertex coloring on the undirected message conflict graph \cite{Yossef_Birk_Jayram_Kol_Trans}, where the clique cover number of the former is equal to the chromatic number of the latter.

\subsection{Relationship between TIM and index coding}
As shown in \cite{Jafar_TIM}, for corresponding instances of the TIM problem and the index coding problem, described by the same topology matrix ${\Tm}$ and message set $\Mc$,  the index coding capacity region is an outer bound on the TIM DoF region, and the two are equivalent under linear solutions \cite{Jafar_TIM}.  Fig. \ref{fig:timic} shows an example.


\subsection{Key Definitions}
Note that some more definitions pertaining to graph theory are given in the Appendix.
\begin{definition}[All-Unicast]
The all-unicast setting means that from each Source $S_i$ there is an independent message $W_{ji}$ to each Destination $D_j$ if $t_{ji}=1$ ($a_{ji}=0$), i.e., if there exists a non-trivial channel between them. Note that this includes arbitrary message sets as special cases, by setting the rates of some messages to zero.
\end{definition}


\begin{definition}[Cycle]
A {cycle} is a set of vertices and edges that form a closed loop. 
The {length} of a cycle is the number of vertices in this cycle. A cycle of length 6 or more is called a long cycle.
\end{definition}

\begin{definition}[Chord]
A {chord} is an edge that connects two non-adjacent vertices of a cycle.
\end{definition}

\begin{definition}[Chordal Bipartite Network]
A TIM (index coding) instance is {referred to as a chordal bipartite network} if its  network topology graph does not contain a chordless long cycle. 
{Chordal bipartite networks are exactly the graph class of chordal bipartite graphs in graph theory.}

\end{definition}

Note that {the network topology graph is always bipartite, and cannot contain odd length cycles. These graphs are simple, so there are no multiple edges and thus no length-2 cycles. In addition,} because a source (destination) is not connected to any other sources (destinations), it is not possible for a length-4 cycle to contain a chord {in the network topology graphs.} 
Thus the chordal bipartite {property in the network topology graph} can be interpreted as -- \emph{any cycle that can contain a chord, must contain a chord.} 

\begin{remark}
Chordal bipartite networks include some special network topology graphs, such as forests and trees, convex/biconvex bipartite graphs, bipartite permutation graphs, bipartite distance hereditary graphs, and difference graphs \cite{PerfectGraph}. 
\end{remark}

\begin{definition}[TDMA \cite{Maleki_Jafar_Convex}]
TDMA refers to a {time-slotted transmission} scheme that schedules messages $\Wc_o$ for transmission simultaneously {in each slot} only if they are non-interfering (orthogonal), i.e., $t_{j'i}  = t_{ji'} = 0$ for TIM ($a_{j'i}  = a_{ji'} = 1$ for index coding),  $\forall ~W_{ji},W_{j'i'} \in \Wc_o,~ i \neq i', j \neq j' \notag$. 
\end{definition}

In the message conflict graph, two adjacent vertices (messages) are interfering, and should be scheduled orthogonally in different slots. If we associate each slot with a different color, TDMA can be done using vertex coloring on the message conflict graphs, where two adjacent vertices (messages) should be assigned  different colors. Associating the slots with a number of different colors, we generalize it to fractional coloring (see the definition in the Appendix). As a result, TDMA is also called fractional coloring in index coding context. 


\begin{definition}[Set of Cliques of Conflict Graph ($\mathcal{Q}$)]
A clique $Q$ of the message conflict graph is a set of vertices (messages) such that any two are adjacent.
The set of all cliques of the message conflict graph is denoted as $\mathcal{Q}$.
\end{definition}

In the rest of this paper, we identify the optimality of TDMA (fractional coloring) through the structural property of the bipartite network topology graphs, and characterize the DoF (capacity) region via vertex coloring on the undirected message conflict graphs. 

\section{Main Result}

Our main result  for the TIM setting is stated in the following theorem.


\begin{theorem}
\label{theorem:chordal}
TDMA achieves the all-unicast DoF region of the TIM problem {if and only if} the network topology is chordal {bipartite}. For chordal {bipartite} networks, the DoF region of the TIM problem is characterized through the cliques $Q\in\mathcal{Q}$ of the message conflict graph, as follows:
\begin{align*}
\forall~Q \in \mathcal{Q}, ~~ \begin{matrix}   
\sum_{W_{ji} \in Q}d_{ji} \le 1
\end{matrix}    
\end{align*}
where $d_{ji}$ is the DoF of the message $W_{ji}$.
\end{theorem}
\begin{proof}
See Section \ref{sec:proof}.
\end{proof}

\begin{remark}
The  DoF region as specified above is comprised of  inequalities corresponding to all possible cliques in the message conflict graph. There may be exponentially many cliques, many of them redundant because they are contained within larger cliques. Therefore,  the representation may be simplified by restricting $\Qc$ to be the set of maximum cliques.
A {maximum clique} is the clique with the maximum possible number of vertices (messages). Any induced subgraph of a maximal clique is still a clique.
Such a simplification is valid because the inequality associated with the maximal clique implies all other inequalities of the cliques with smaller size.
Since our focus in this paper is on the DoF region characterization, computation and complexity issues of listing all maximal cliques are beyond the scope of this paper.
\end{remark}

Next we illustrate this result through some examples. 

\begin{example}
Consider the one-dimensional  TIM network instance with message $W_{ii}$ from $S_i$ to $D_i$ as shown in Fig.~\ref{fig:IC_ex}, where convexity applies to the sources only. {Note here that we only consider a subset of messages $\Mc=\{W_{11},W_{22},W_{33},W_{44},W_{55}\}$ rather than the all-unicast setting, so we construct the message conflict graph only with  respect to $\Mc$.} Because the network topology graph is chordal {bipartite}, Theorem 1 applies and TDMA achieves the DoF region. {Note that the 6-vertex cycle $S_1 - D_2 - S_3 - D_3 - S_5 - D_4 - S_1$ is not chordless, because it indeed contains a chord $S_1-D_3$.} Considering all sets of conflicting messages, i.e., all cliques of the conflict graph shown in Fig. \ref{fig:IC_ex}, {and removing the redundant ones}, we have the DoF region
\begin{align*}
\Dc=\left\{ (d_{11},\dots,d_{55}) \in \mathbb{R}_+^5 ~ \bigg|  \Pmatrix{d_{11}+d_{22} +d_{33} \le 1 \\ d_{11}+d_{44} \le 1\\ d_{33}+d_{55} \le 1 \\ d_{44}+d_{55} \le 1} \right\}
\end{align*}
As said, this DoF region can be simply obtained by listing all the maximal cliques, i.e., the message sets $\{W_{11},W_{22},W_{33}\}$, $\{W_{11},W_{44}\}$, $\{W_{33},W_{55}\}$, and $\{W_{44},W_{55}\}$.
\end{example}

\begin{example}
Consider the TIM instance in Fig.~\ref{fig:chordal_ex}, with  message set $\Mc = \{W_{13},W_{24},W_{31},W_{45},W_{52},W_{55},W_{73}\}$ (which is a subset of the all-unicast message set). Since the network is chordal {bipartite}, Theorem~\ref{theorem:chordal} applies and we have the DoF region {by listing all the inequalities associated with the maximal cliques, where the constraints of smaller cliques are automatically implied.}
\begin{align*}
\Dc=\left\{ d_{\Mc} \in \mathbb{R}_+^7 ~ \Bigg|  \Pmatrix{d_{24}+d_{31} \le 1
\\  d_{31}+d_{13} + d_{73} \le 1 \\ d_{13}+d_{45} + d_{73} \le 1 \\ d_{45}+d_{55} + d_{73} \le 1 \\ d_{52}+d_{45} + d_{55} \le 1} \right\}
\end{align*}
\end{example}

Additionally, we are able to show (proof relegated to Section \ref{sec:one-d})
that one-dimensional TIM instances with either source convexity or  destination convexity (not necessarily both) are chordal {bipartite}. Thus, the optimality of TDMA shown in \cite{Maleki_Jafar_Convex} for one-dimensional convex networks, applies with either source \emph{or} destination convexity, to any arbitrary unicast message set, and  to the entire DoF region.


We conclude this section by translating the result for TIM to the index coding setting, as stated in the following theorem.

\begin{theorem}\label{thm:index}
Fractional coloring  achieves the all-unicast capacity region of the index coding problem associated with a network topology graph {if and only if} the network topology is chordal bipartite. For chordal bipartite networks, the capacity region of the index coding problem is the set of rate tuples comprised of $R_{ji}$ in $\mathbb{R}^+$ that satisfy the following constraints:
\begin{align*}
\forall~Q \in \mathcal{Q}, ~~ \begin{matrix}   \sum_{W_{ji} \in Q} R_{ji} \le 1
\end{matrix}    
\end{align*}
where $R_{ji}$ is the rate of the message $W_{ji}$.
\end{theorem}

\begin{proof}
See Section \ref{sec:proof}.
\end{proof}

\section{Conclusion}
We show that the necessary and sufficient condition for TDMA (fractional coloring) to achieve the all-unicast DoF (capacity) region of TIM (index coding) is that the network topology should be chordal {bipartite}. The absence/presence of chordless cycles prevents/creates opportunities for more sophisticated achievable schemes. 
Among other interesting observations, we note that for chordal bipartite networks while fractional coloring is needed to achieve the capacity region, {the conventional (non-fractional)} coloring suffices to achieve all the corner points, {i.e., the extreme points of the capacity region polytope}. Finally, in the TIM context, 
the main results hold regardless of channel coherence time. The proof is presented in Section \ref{sec:time-varying}.

\section{Proofs}
\label{section:proof}
\subsection{Graph Theoretic Preliminaries} 
{
We assume the readers will have some basic knowledge of graph theory. For the readers who are not familiar with the graph theory notions in use here, we refer to the Appendix for additional definitions.
}

Consider the undirected topology graph $\Gc$, {which is bipartite}. 
The {\em line graph} of $\Gc$ is another graph, denoted by $\Gc_e$, such that each vertex of $\Gc_e$ represents an edge of $\Gc$, and any two vertices in $\Gc_e$ are adjacent if and only if their corresponding edges in $\Gc$ have an endpoint in common.
The {\em square} of a graph $\Gc$, denoted by $\Gc^2$, is another graph 
that has the same set of vertices in $\Gc$, but in which two vertices are adjacent when their distance in $\Gc$ is at most 2. The {\em distance} between two vertices is the number of edges in a shortest path connecting them.
Interestingly, $\Gc_e^2$ is the message conflict graph corresponding to the network topology graph $\Gc$. We state this as the following lemma.
The proof is presented in Section \ref{sec:conflict_graphs}.

\begin{lemma}
\label{lemma:conflict_graphs}
For any network topology graph 
$\Gc$, the square of its line graph, 
$\Gc_e^2$, is its message conflict graph.
\end{lemma}

Next, we carry over the chordal bipartite property of $\Gc$ to the perfect-graph property of 
$\Gc_e^2$. $\Gc$ is a chordal bipartite graph (see Chapter 12.4 in \cite{PerfectGraph}), thus weakly chordal \cite{ChordalBipartite}. It is proved in \cite{ChordalBipartite} that if $\Gc$ is weakly chordal, then $\Gc_e^2$ is also weakly chordal. As weakly chordal graphs are subclass of perfect graphs (see Chapter 66.5d in \cite{Schrijver}), $\Gc_e^2$ is perfect. We state this crucial result as the following lemma.

\begin{lemma}
\label{lemma:perfect}
If $\Gc$ is chordal {bipartite}, then $\Gc_e^2$ is perfect.
\end{lemma}

\begin{remark}
When a subset of messages is considered, we have an induced subgraph of $\Gc_e^2$ as the message conflict graph. Let $\Mc$ be the message set of interest. The message conflict graph with respect to the message set $\Mc$ can be represented by the induced subgraph $\Gc_e^2[\Mc]$. The chordal bipartite, weakly chordal, and perfect properties are also inherited by the induced subgraphs \cite{PerfectGraph}.
\end{remark}

Let us also recall the definition of a Demand Graph \cite{Jafar_TIM, Neely_Tehrani_Zhang_Trans}. 

\begin{definition}[Demand Graph]
 The {demand graph} is a directed bipartite graph with messages on one side and destinations on the other, with a directed edge from a message to a destination if this message is intended for this destination, and with a directed edge from a destination to a message if the source from which this message originates is not connected  to the destination in the network topology graph.
\end{definition}

\begin{lemma}[From \cite{Jafar_TIM, Neely_Tehrani_Zhang_Trans, Yossef_Birk_Jayram_Kol_Trans}] \label{lemma:acyclic}
For index coding, the sum-capacity of a message set that forms an acyclic demand graph  is bounded by 1.
\end{lemma}

For a chordal bipartite network, an acyclic demand graph connects to a clique in $\Gc_e^2$, as stated in the following lemma. The proof is presented in Section \ref{sec:cliquebound}.

\begin{lemma} \label{lemma:cliquebound}
If $\Gc$ is chordal {bipartite}, then for each clique in $\Gc_e^2$, the associated messages form an acyclic demand graph.
\end{lemma}

{\color{black}
\subsection{Proofs of Theorem \ref{theorem:chordal} and Theorem \ref{thm:index}} \label{sec:proof}
}

As Theorem \ref{theorem:chordal} and Theorem \ref{thm:index} have similar forms, we present their proofs simultaneously. 

{\bf Sufficiency:} We prove that if a network is chordal {bipartite}, then TDMA (fractional coloring) achieves the all-unicast DoF (capacity) region.

First consider the outer bound. It suffices to prove the outer bound of the index coding problem (Theorem \ref{thm:index}) because the capacity region of an index coding instance is an outer bound on the  DoF region for the associated TIM instance \cite{Jafar_TIM}. Therefore the outer bound of the TIM problem (Theorem \ref{theorem:chordal}) is implied by that of the index coding problem. 

We now prove the outer bound of the index coding problem.
As Lemma \ref{lemma:cliquebound} shows that each clique in $\Gc_e^2$ corresponds to an acyclic demand graph, and by Lemma \ref{lemma:acyclic} the sum capacity 
of associated messages in an acyclic demand graph is bounded by 1, we obtain the following outer bounds.
\begin{align*}
\forall~Q \in \mathcal{Q}, ~~ \begin{matrix}  
\sum_{W_{ji} \in Q} R_{ji} \le 1, 
\end{matrix}    
\end{align*}
The above inequalities are called the clique inequalities \cite{Schrijver}. 

Next we proceed to the achievability. It suffices to prove the achievability for the TIM problem (Theorem \ref{theorem:chordal}) because any achievable DoF tuple of a TIM instance translates to the same achievable rate tuple of the associated index coding instance \cite{Jafar_TIM}. Therefore the achievability of the index coding problem (Theorem \ref{thm:index}) is implied by that of the TIM problem. We now prove the achievability of the TIM problem.

We show that the following outer bound DoF region is achievable by TDMA. 
\begin{align*}
\forall~Q \in \mathcal{Q}, ~~ \begin{matrix}  
\sum_{W_{ji} \in Q}d_{ji} \le 1
\end{matrix}    
\end{align*}
To this end, we show that the outer bound region has integral vertices, meaning that each coordinate of the DoF tuples of the vertices is either 0 or 1. Note that an integral vertex of the polytope defined by the clique inequalities corresponds to a set of orthogonal messages that do not conflict, such that they can be scheduled over one time slot. As each vertex of the outer bound region can be achieved by one shot scheduling, time sharing between these vertices can achieve the whole region and the overall scheme is TDMA.

We are left to prove the outer bound region has integral vertices. This follows from the fact that $\Gc_e^2$ is a perfect graph and it is known that the polytope defined by the clique inequalities of a perfect graph has integral vertices (see \cite{chvatal} and Chapter 65 in \cite{Schrijver}).

The sufficiency proof is complete.

{\bf Necessity:}
{The necessity proofs for Theorem \ref{theorem:chordal} and Theorem \ref{thm:index} are identical, i.e., we only need to show that there exists at least one unicast message setting for which TDMA (fractional coloring) is suboptimal. Specifically, }we want to show that if a network is not chordal bipartite (contains chordless cycles with length $2n, n = 3, 4, \ldots$), then TDMA (fractional coloring) does not achieve the DoF (capacity) region {for at least one message setting}. Let us start with the TIM setting (Theorem \ref{theorem:chordal}).

Suppose now the network topology graph contains a chordless cycle with $2n$ vertices, which consist of $n$ distinct source nodes, labeled $S'_1, \ldots, S'_n$ and $n$ distinct destination nodes, labeled $D'_1,\ldots,D'_n$. We consider the sub-network induced by these nodes. As the cycle is chordless, the sub-network topology is cyclic where each Source $S'_i, i \in \{1,\ldots,n\}$ is connected to two destinations $D'_{i-1}, D'_{i}$ and each Destination $D'_i$ is connected to two sources $S'_i, S'_{i+1}$ (source/destination indices are interpreted modulo-$n$, i.e., $n+1 = 1, 0 = n$). An example with $n = 4$ is shown in Fig. \ref{fig:timic}.
To prove the desired claim, it suffices to find a DoF tuple that is achievable, thus inside the DoF region, but cannot be achieved by TDMA.

We consider the cases where $n$ is odd or even separately.

When $n$ is odd, we consider the interference channel message setting, i.e., there are $n$ desired messages in the sub-network, one each from Source $S'_i$ to Destination $D'_i$. As each destination only suffers interference from one non-desired source, CDMA (i.e., multicast) can achieve DoF tuple $(1/2, 1/2, \ldots, 1/2)$ with $n$ elements \cite{Jafar_TIM}. This is because over two channel uses, each destination sees two independent linear equations in the two symbols (one desired and one interfering) that it is able to hear, from which it can resolve both. However, it is easily seen that TDMA cannot schedule more than $(n-1)/2 < n/2$ messages over one time slot, and is therefore unable to achieve this DoF tuple.

When $n$ is even, we consider the all-unicast message setting and the DoF tuple $(1/3, 1/3, \ldots, 1/3)$ with $2n$ elements, which is achievable by interference alignment \cite{Jafar_TIM}, but not by  TDMA. TDMA cannot schedule more than $n/2 < 2n/3$ messages at the same time, e.g., if we schedule the message from $S'_i$ to $D'_i$, then messages from $S'_i$ to $D'_{i-1}$, from $S'_{i+1}$ to $D'_i$ and $S'_{i-1}$ to $D'_{i-1}$ cannot be scheduled. Thus, TDMA is again sub-optimal.

This completes the necessity proof for the TIM setting (Theorem \ref{theorem:chordal}). We now consider the necessity proof for the index coding setting (Theorem \ref{thm:index}). The same proof applies by noting that the above cases only use linear schemes and index coding and TIM are equivalent under linear schemes \cite{Jafar_TIM}.

\begin{remark}
The prerequisite of the necessity is under the all-unicast setting, where all messages are taken into account. There exist some topologies that are not chordal bipartite, and TDMA still achieve the DoF region for a particular subset of messages. This does not contradict the necessity.
\end{remark}

\begin{remark}
From the DoF region, it is not hard to verify that symmetric DoF value is given by $1/\chi(\Gc_e^2)$, where $\chi(\Gc_e^2)$ is the chromatic number of $\Gc_e^2$, and sum DoF value is given by the independence set number of $\Gc_e^2$.
\end{remark}

\subsection{Proof of Lemma \ref{lemma:conflict_graphs}} \label{sec:conflict_graphs}
First, $\Gc_e^2$ and the message conflict graph of $\Gc$ have the same vertex set. In the message conflict graph, there is a vertex for each message in $\Gc$. In $\Gc_e^2$, there is a vertex for each edge in $\Gc$ and in the all-unicast message setting, each edge in $\Gc$ corresponds to a message. Thus the claim follows.

Second, we prove $\Gc_e^2$ and the message conflict graph of $\Gc$ have the same edge set. 
In the message conflict graph, two messages (vertices) $W_{ji}, W_{j'i'}$ are connected if and only if they originate from the same source ($i = i'$), or are intended for the same destination ($j = j'$), or one source interferes with the other destination ($a_{j'i} = 0 (t_{j'i} = 1)$ or $a_{ji'} = 0 (t_{ji'} = 1)$). When $i = i'$ or $j = j'$, the two edges representing $W_{ji}, W_{j'i'}$ in $\Gc$ share a common vertex such that these two messages (vertices) are connected in $\Gc_e$ (have distance 1, thus connected in $\Gc_e^2$). When $a_{j'i} = 0 (t_{j'i} = 1)$ or $a_{ji'} = 0 (t_{ji'} = 1)$, the two edges representing messages $W_{ji}, W_{j'i'}$ in $\Gc$ both connect to the edge representing message $W_{j'i}$ or $W_{ji'}$ such that these two messages (vertices) are both connected to message (vertex) $W_{j'i}$ or $W_{ji'}$ in $\Gc_e$ (have distance 2, thus connected in $\Gc_e^2$). Conversely, whenever an edge exists between two messages (vertices) in $\Gc_e^2$ (have distance 1 or 2 in $\Gc_e$), the messages conflict in $\Gc$ (thus connected in the message conflict graph). Then we have the desired claim.

Therefore, $\Gc_e^2$ is the message conflict graph of $\Gc$. Note that this result holds regardless of whether $\Gc$ is chordal {bipartite} or not.

\subsection{Proof of Lemma \ref{lemma:cliquebound}} \label{sec:cliquebound}
We show that if a set of messages forms a clique in $\Gc_e^2$ (they mutually conflict), the demand graph formed by these messages and their desired destinations must be acyclic, given $\Gc$ is chordal bipartite. To set up a proof by contradiction, suppose a set of messages forms a clique in $\Gc_e^2$ and {\color{black} there exists a directed cycle (di-cycle) in the induced demand graph comprised of these messages and their desired destinations.  Let $\mathcal{G}_c$ be the shortest such di-cycle. $\mathcal{G}_c$ must be a chordless di-cycle because if $\mathcal{G}_c$ contains a chord, then the chord splits $\mathcal{G}_c$ into two cycles. One of the two cycles is a di-cycle and it is shorter than $\mathcal{G}_c$, contradicting the assumption that $\mathcal{G}_c$ is the shortest.
The length of $\mathcal{G}_c$ is denoted as $k$.

We argue that $\mathcal{G}_c$ cannot contain vertices corresponding to two or more messages that originate at the same source node. To see this, let us assume the opposite, that $\mathcal{G}_c$ contains vertices corresponding to two messages (denoted as $W_i', W_j'$) that originate from the same source (denoted as $S^*$). Since $\mathcal{G}_c$ is a di-cycle, there must be an arc from a destination in $\mathcal{G}_c$ (say $D_l'$) to $W_i'$. From the definition of the demand graph, we know that $D_l'$ is not connected to $S^*$ in the network topology graph. Since $W_j'$ also originates at $S^*$, there must be an arc from $D_l'$ to $W_j'$ in the demand graph as well. Thus, we have two outgoing arcs in $\mathcal{G}_c$ from $D_l'$, to $W_i'$ and $W_j'$. But this is not possible because a node in a chordless di-cycle  cannot have two outgoing arcs. Therefore, we have proved that $\mathcal{G}_c$ cannot contain vertices corresponding to two or more messages that originate at the same source node. Similarly, $\mathcal{G}_c$ cannot contain vertices corresponding to two or more messages that are intended for the same destination node. 

Next we proceed to show the contradiction, that the demand graph formed by a set of messages that forms a clique in $\Gc_e^2$ can not contain a di-cycle. 
First, because the demand graph is bipartite, $k$ must be even such that $k = 2n, n\in\mathbb{N}$.

Second, $n \neq 1$ because such a length-$2$ di-cycle in the demand graph means that a destination wants a message and is not connected 
to the source that emits the message in the network topology graph
, which is not possible. 

Third, $n \neq 2$, because otherwise $\mathcal{G}_c$ contains two messages, $W_i', W_j'$ (which originate at sources $S_i'$ and $S_j'$), and their desired destinations, $D_i', D_j'$, respectively, such that in the network topology graph $D_i'$ is not connected to $S_j'$, and $D_j'$ is not connected to $S_i'$. Therefore, messages $W_i', W_j'$ do not conflict. This contradicts the assumption that these two messages form a clique in  $\Gc_e^2$.

Finally, we consider $n = 3, 4, \ldots$. 
Since the sources from which the messages originate are distinct, let us replace each message node in $\mathcal{G}_c$ with the source node at which it originates. With this substitution, $\mathcal{G}_c$ is made up of $n$ sources, denoted as $S_1', \cdots, S_n'$ and $n$ destinations, denoted as $D_1', \cdots, D_n'$. Without loss of generality, let us assume that in this chordless di-cycle, there is an arc from $S_i'$ to $D_i', i \in \{1,\cdots,n\}$ and an arc from $D_i'$ to $S_{i+1}'$, where the indices are interpreted modulo-$n$. As the di-cycle is chordless, it  contains only these $2n$ arcs. Now  compare $\mathcal{G}_c$ with the (undirected) network topology graph (denoted as $\Gc^*$) induced by sources $S_1',\cdots, S_n'$ and destinations $D_1', \cdots, D_n'$. In $\Gc^*$, direct links (edges between $S_i'$ and $D_i'$) remain and cross links are the complements of those in  $\mathcal{G}_c$, i.e., $D_{i}'$ is connected to $S_{i+2}', S_{i+3}', \cdots, S_{i+N-1}'$. This is because there is an arc from Destination $D_i'$ to message $W_j', j \neq i$ in the demand graph if and only if $D_i'$ is not connected to Source $S_j'$ that emits $W_j'$ in $\Gc^*$.
$\Gc^*$ contains a  cycle of length 6, $S_1' - D_1' - S_3' - D_n' - S_2' - D_2' - S_1'$. This cycle is chordless because $D_1'$ is not connected to $S_2'$; $D_n'$ is not connected to $S_1'$; and $D_2'$ is not connected to $S_3'$ in the network topology graph.
 This contradicts the assumption that the overall network topology graph $\Gc$ (which contains $\Gc^*$ as a subgraph) does not contain chordless cycles with length 6 or more.

Therefore, for chordal bipartite networks $\Gc$, whenever we have a clique in $\Gc_e^2$, the associated messages form an acyclic demand graph. The proof is complete.

\begin{remark}
Note that a clique in $\Gc_e^2$ may not correspond to an acyclic demand graph if $\Gc$ is not chordal bipartite. For a counterexample, consider a cyclic network with 3 sources, labeled $S'_1, S'_2, S'_3$ and 3 destinations, labeled $D'_1, D'_2, D'_3$. In the network topology graph, $S'_1$ is connected to $D'_3, D'_1$, $S'_2$ is connected to $D'_1, D'_2$, and $S'_3$ is connected to $D'_2, D'_3$. The network topology graph is not chordal bipartite as all vertices form a length-6 chordless cycle. Consider the messages $W'_{11}, W'_{22}, W'_{33}$. They mutually conflict and form a clique in $\Gc_e^2$. But the demand graph formed by them is not acyclic as CDMA can achieve DoF $1/2$ per message and the sum DoF is not bounded up by 1.
\end{remark}

\subsection{One-dimensional Convex Networks} \label{sec:one-d}

We show that one-dimensional TIM instances with either source convexity or destination convexity (not necessarily both) are chordal bipartite.

We first define convexity. In a one-dimensional network, the source and destination nodes are placed along a straight line. We define the relation $a < b$ between two nodes to indicate that node $a$ is ``to the left of" node $b$.

\begin{definition} [Source Convexity]
Source convexity refers to the property that if a source (say $S_i$) can be heard by two destination nodes (say $D_j, D_k$),
then it must also be heard by all other destination nodes that are in between, i.e., $(D_j < D_l < D_k)  ~ \mbox{AND} ~ (t_{ki} = t_{ji} = 1) \Rightarrow t_{li} = 1$. 
\end{definition}

\begin{definition} [Destination Convexity]
Destination convexity refers to the property that if a destination (say $D_i$) can hear two source nodes (say $S_j, S_k$), then it must also hear all other source nodes that are in between, i.e., $(S_j < S_l < S_k)  ~ \mbox{AND} ~ (t_{ij} = t_{ik} = 1) \Rightarrow t_{il} = 1$. 
\end{definition}

We next proceed to the proof. Since the name of source or destination is entirely cosmetic, we consider a one-dimensional TIM instance with only source convexity, without loss of generality. Similar proof applies to cases with only destination convexity as well.
To set up a proof by contradiction, suppose the TIM instance is not chordal bipartite, i.e., its network topology graph contains a chordless cycle with length $2n, n = 3, 4, \ldots$, which corresponds to a cyclic sub-network with $n$ distinct sources, labeled $S'_1, \ldots, S'_n$ and $n$ distinct destinations, labeled $D'_1,\ldots,D'_n$. Source $S'_i, i \in \{1,\ldots,n\}$ is connected to two destinations $D'_{i-1}, D'_{i}$ and Destination $D'_i$ is connected to two sources $S'_i, S'_{i+1}$ (source/destination indices are interpreted modulo-$n$, i.e., $n+1 = 1, 0 = n$). As the cycle is chordless, each source is connected to only two destinations. Further, because of source convexity, the destinations that are connected to the same source must be consecutive. For example, as $S'_2$ is connected to only $D'_1$ and $D'_2$, there can not be any destination in the interval between $D'_1$ and $D'_2$. Similarly, there is no destination in between $D'_i$ and $D'_{i+1}$, such that the order of the destinations in one straight line must appear as $D'_{1} < D'_{2} < \cdots < D'_{n}$ or $D'_{1} > D'_{2} > \cdots > D'_{n}$. In both cases,  $D'_{1}$ and $D'_{n}$ are not consecutive. We arrive at a contradiction.

\subsection{Coherence Time} \label{sec:time-varying}
We show that Theorem \ref{theorem:chordal} holds regardless of channel coherence time, i.e., TDMA achieves the all-unicast DoF region of the TIM problem if and only if the network is chordal bipartite, regardless of channel coherence time.

We now do not require the channel coefficients $h_{ji}(t)$ to be constant. Instead, $h_{ji}(t)$ can vary in an arbitrary manner as long as the values are bounded away from zero and infinity, i.e., there is no requirement on the channel coherence time.

{\bf Sufficiency:} We prove that if a TIM network is chordal bipartite, then TDMA achieves the all-unicast DoF region, regardless of channel coherence time. 

The outer bounds provided in Section \ref{sec:proof} hold regardless of channel coherence time.
Also, TDMA scheme applies to both constant and time-varying channel setting, such that the achievability proof in Section \ref{sec:proof} is not affected. Hence sufficiency is proved.

{\bf Necessity:} We prove that if a TIM network is not chordal bipartite, then TDMA does not achieve the all-unicast DoF region, regardless of channel coherence time.

We follow the proof in Section \ref{sec:proof}. Suppose the network topology graph is not chordal bipartite, such that it contains a chordless length-$2n$ cycle, $n = 3, 4, \ldots$. Such a chordless cycle corresponds to a cyclic sub-network. We show that TDMA can not achieve the DoF region for such a cyclic sub-network (and therefore the all-unicast DoF region).

We also consider the cases where $n$ is odd or even separately.

When $n$ is odd, we use the same interference channel message setting and consider DoF tuple $(1/2, 1/2, \ldots, 1/2)$. As CDMA can be applied to time-varying channels, the DoF tuple is still achievable, regardless of channel coherence time. However, TDMA can not schedule more than $(n-1)/2$ messages over one time slot, and is therefore unable to achieve this DoF tuple.

When $n$ is even, we still consider the all-unicast message setting. In this case, the DoF tuple $(1/3, 1/3, \ldots, 1/3)$ does not work as the interference alignment scheme used to achieve this tuple requires the channels to be constant for 3 symbol periods \cite{Jafar_TIM}. Instead, we consider the sum-DoF value. As shown in Section \ref{sec:proof}, TDMA can not achieve more than $n/2$ sum-DoF. However, the scheme in \cite{CBIA} can achieve $(n+1)/2$ sum-DoF, regardless of channel coherence time. See Fig. 9 in \cite{CBIA} for a pictorial illustration for $n=4$ case. Thus, TDMA is again sub-optimal, regardless of channel coherence time. This completes the necessity proof.

\appendix
Throughout this paper, we consider simple graphs, which are unweighted, containing no loops for a single vertex or multiple edges between two vertices.

%
%
A subgraph of $\Gc=(\Vc,\Ec)$ containing a subset of vertices $\Sc$ $(\Sc \subseteq \Vc)$ is said to be an {\em induced subgraph}, denoted by $\Gc[\Sc]$, if for any pair of vertices $u$ and $v$ in $\Sc$, $uv$ is an edge of $\Gc[\Sc]$ if and only if $uv$ is an edge of $\Gc$.

The {\em chromatic number} of $\Gc$, denoted by $\chi(\Gc)$, is the smallest number of colors that can be assigned to the vertices of $\Gc$, such that no two adjacent vertices have the same color. Such an assignment method is referred to ordinary {\em graph coloring}.
A graph $\Gc$ is said to be {\em $n:m$-colorable} if each vertex in $\Gc$ can be assigned a set of $m$ distinct colors in which the colors are drawn from a palette of $n$ colors, such that any adjacent vertices have no colors in common. This coloring method is also referred to as {\em fractional coloring}, which is a generalization of ordinary graph coloring.
Denote by $\chi_m(\Gc)$ the minimum required number, $n$, such that the {\em fractional chromatic number} $\chi_f(\Gc)$ can be
defined as
\begin{align}
\chi_f(\Gc) = \lim_{m \to \infty} \frac{\chi_m(\Gc)}{m} = \inf_m \frac{\chi_m(\Gc)}{m}.
\end{align}
A {clique} is a subgraph of a graph $\Gc$ such that any two vertices in this subgraph are adjacent. The size of a clique is the number of vertices. A {maximum clique} is the  clique with the maximum possible size in $\Gc$. The {\em clique number} of $\Gc$, denoted by $\omega(\Gc)$, is the number of vertices in the maximum clique.  $\chi(\Gc) \ge \omega(\Gc)$ for any undirected graph.
The {independent set} of a graph $\Gc$ is a set of vertices such that any two vertices are not adjacent. The {\em independent set number}, denoted by $\alpha(\Gc)$, is the cardinality of the largest independent set.
%

%
A {\em perfect graph} is a graph $\Gc$ in which the chromatic number of every induced subgraph $\Hc$ of $\Gc$ equals  the clique number of this subgraph, i.e., $\chi(\Hc) = \omega(\Hc)$.
A {chordless cycle} is a cycle with no edges between any non-consecutive vertices. The length of a cycle is the number of vertices in this cycle.
A {\em hole} is a chordless cycle with five or more vertices, and an {\em antihole} is the complement of a hole. The {\em complement} of a graph, is another graph containing the same vertices set, but an edge in one graph if and only if it is not in the complement. The underlying undirected graph of a directed graph is obtained by ignoring the direction in the directed graph.
An undirected graph is {\em perfect} if and only if it  contains neither odd holes nor odd antiholes as induced subgraphs. The odd hole is a hole with odd length, and the odd antihole is its complement.
A chordless directed cycle (di-cycle) refers to the induced sub-digraph with $n$ nodes $\{v_0, v_1,\cdots, v_{n-1}\}$ and arcs $\{(v_0,v_1),(v_1, v_2),\cdots, (v_{n-2}, v_{n-1}), (v_{n-1},v_0) \}$, beyond which there do not exist any other arcs. 
A directed graph (or subgraph) is acyclic if it does not contain any di-cycles.

A graph is {\em chordal} (or triangulated) if there is no induced subgraph with chordless cycles of length greater than three, i.e., every cycle with length greater than three has a chord. 
A graph is {\em weakly chordal} (or weakly triangulated) if it is hole-free and antihole-free in its induced subgraphs.
Both chordal and weakly chordal graphs are subclasses of perfect graphs. 

A graph is {\em chordal bipartite} if it is an undirected bipartite graph and there is no induced subgraph with chordless cycles of length greater than four, i.e., every cycle of length at least six has a chord.
The chordal bipartite graph is both bipartite and chordal, and hence weakly chordal.
For a graph $\Gc$, if $\Gc$ is weakly chordal, then $\Gc_e^2$ is also weakly chordal \cite{ChordalBipartite}, and hence perfect.



\end{document}